# An Alternative Approach of Steganography using Reference Image

Samir Kumar Bandyopadhyay
Senior Member IEEE
Dept. of Computer Science & Engineering,
University of Calcutta, India
Email: skb1@vsnl.com

Indra Kanta Maitra
B.P. Poddar Institute of Engineering and Management,
Kolkata, India
Email: ikm.1975@vsnl.net

**Abstract**

This paper is to create a practical steganographic implementation for 4-bit images. The proposed technique converts 4 bit image into 4 shaded Gray Scale image. This image will be act as reference image to hide the text. Using this grey scale reference image any text can be hidden. Single character of a text can be represented by 8-bit. The 8-bit character can be split into 4X2 bit information. If the reference image and the data file are transmitted through network separately, we can achieve the effect of Steganography. Here the image is not at all distorted because said image is only used for referencing. Any huge amount of text material can be hidden using a very small image. Decipher the text is not possible intercepting the image or data file separately. So, it is more secure.

## 1. Introduction

The growth of high-speed computer networks and that of the Internet, in particular, has increased the ease of Information Communication. Ironically, the cause for the development is also of the apprehension use of digital formatted data. In comparison with Analog media, Digital media offers several distinct advantages but this type advancement in the field of data communication in other sense has hiked the fear of getting the data intercepted at the time of sending it from the sender to the receiver.

Cryptography was created as a technique for securing the secrecy of communication and many different methods have been developed to encrypt and decrypt data in order to keep the message secret. Unfortunately it is sometimes not enough to keep the contents of a message secret, it may also be necessary to keep the existence of the message secret. The technique used to implement is called steganography. It is the art and science of invisible communication. This is accomplished through hiding information in other information, thus hiding the existence of the communicated information.

The aim of this paper is to create a practical steganographic implementation for 4-bit images. The proposed method converts 4 bit image into 4 shaded Gray Scale image. This image will be act as reference image to hide the text.

## 2. Steganography

The word steganography is originally derived from Greek words, which mean "Covered Writing". It has been used in various forms for thousands of years. In the 5th century BC Histaiacus shaved a slave's head, tattooed a message on his skull and the slave was dispatched with the message after his hair grew back [1, 2, 3, 4]. In Saudi Arabia at the King Abdulaziz City of science and technology, a project was initiated to translate into English some ancient Arabic manuscripts on secret writing which are believed to have been written 1200 years ago. Some of these manuscripts were found in Turkey and Germany [5]. Five hundred years ago, the Italian mathematician Jérôme Cardan reinvented a Chinese ancient method of secret writing. The scenario goes as follows: a paper mask with holes is shared among two parties, this mask is placed over a blank paper and the sender writes his secret message through the holes then takes the mask off and fills the blanks so that the message appears as an innocuous text. This method is credited to Cardan and is called Cardan Grille [4].

### *2.1 Digital Steganography*

With the boost in computer power, the internet and with the development of digital signal processing (DSP), information theory and coding theory, steganography has gone "digital". In the realm of this digital world steganography has created an atmosphere of corporate vigilance that has spawned various interesting applications, thus its continuing evolution is guaranteed. Contemporary information hiding is due to [6]. One of the earliest methods to discuss digital steganography is credited to Kurak and McHugh [7], who proposed a method, which resembles embedding into the 4 LSBs (least significant bits). They examined image downgrading and contamination, which is known now as image-based steganography. Information can be hidden inside a multimedia object using many suitable techniques. As a cover object, we can select image, audio or video file. Depending on the type of the cover object, definite and appropriate technique is followed in order to obtain security. In this paper, we will discuss techniques or methods by which we can hide text information in data file using a reference image instead of cover image.

### *2.2 Bitmap Image*

A bitmap image consists of a two-dimensional array of numbers. The color shade displayed for a given picture element (pixel) depends on the number stored in the array for that pixel. The simplest type of image data is black and white. It is a binary image since each pixel is either 0 or 1. The next, more complex type of image data is grey scale, where each pixel takes on a value between zero and the number of gray scales or grey. These images appear like common black and white image – they are black, white, and shades of grey. Most grey scale images today have 256 shades of grey. The most complex type of image is color. Color images are similar to grey scale except that there are three bands or channels, corresponding to the colors red, green, and blue, commonly called RGB image. Thus, each pixel has three values associated with it. The range of color is depend on color depth or bit depth, is a term describing the number of bits used to represent the color of a single pixel in a bitmapped image.

This concept is also known as bits per pixel (bpp). Color depth is only one aspect of color representation expressing how finely levels of color can be expressed; the other aspect is how broad a range of colors can be expressed. The RGB color model, cannot express many colors, notably saturated colors [8, 9,10].

## *2.3 Indexed Color*

In uncompressed BMP files and many other bitmap file formats, image pixels are stored with a color depth of 1, 4, 8, 16, 24, or 32 bits per pixel. Images of 8 bits and fewer can be either greyscale or indexed color. Depending on the color depth, a pixel in the picture will occupy at least n/8 bytes where n is the bit depth, since 1 byte equals 8 bits. Indexed colors image that enables 8 bits per pixel or less to look almost as good as 24 bits per pixel[11, 12,13,14].

The technique determines the 256 most frequently used colors in the image and creates a color lookup table, also called a "color map" or "color palette," that is stored with the image [15,16]. Rather than each pixel in the image having all three RGB colors (one 8-bit red, one 8-bit green and one 8-bit blue), each pixel contains one 8-bit number that indexes into the 256-color lookup table, which contains the RGB values. This is reducing images to their smallest size and these images are most commonly used on Web pages, as they are small and quick to load. The 256-color palette is mapped for best results on the Internet [17].

## 3. Proposed Technique

After reviewing the current papers on the market for steganography, it was determined that there was not a practical implementation for 4-bit images. Although network speed is increasing, and bandwidth problems are decreasing, file size is still of utmost importance and smaller file sizes are optimal in network communication. Thus, the current steganographic use of 24-bit / 8-bit images leads to slower communication and development of a 4-bit image format would be beneficial.

## *3.1 Selection and Grey Conversion*

The aim of this research is to create a practical steganographic implementation for 4-bit images. So, 4-bit is necessary to preserve the information of a single pixel in color bitmap image.

Despite varying bit depths in CGA's graphics mode, CGA processes colors in its palette in four bits, yielding $2^4 = 16$ different colors. The four color bits are arranged according to the RGBI (Red, Green, Blue and Intensity) color model i.e. the lower three bits represent red, green and blue color components; a fourth "intensifier" bit increases the brightness of all three red, green and blue components. In 4-bit color scheme, color number and color return code is not same. Conversion method of color number and its return code is displayed in Table 1.

The color number can be converted to return code by exchanging $1^{st}$ bit with the $3^{rd}$ bit of binary code. The algorithm for aforesaid conversion is given in figure 1.

Table 1: Color Table for Bitmap Image

| Colors | Number (Binary) | Return Code (Binary) |
|---|---|---|
| Black | 0 (0000) | 0 (0000) |
| Red | 1 (0001) | 4 (0100) |
| Green | 2 (0010) | 2 (0010) |
| Brown | 3 (0011) | 6 (0110) |
| Blue | 4 (0100) | 1 (0001) |
| Magenta | 5 (0101) | 5 (0101) |
| Cyan | 6 (0110) | 3 (0011) |
| Light Gray | 7 (0111) | 7 (0111) |
| Dark Gray | 8 (1000) | 8 (1000) |
| Light Red | 9 (1001) | 12 (1100) |
| Light Green | 10 (1010) | 10 (1010) |
| Yellow | 11 (1011) | 14 (1110) |
| Light Blue | 12 (1100) | 9 (1001) |
| Light Magenta | 13 (1101) | 13 (1101) |
| Light Cyan | 14 (1110) | 11 (1011) |
| White | 15 (1111) | 15 (1111) |

*Algorithm 1 Conversion*
*Input: Color Number*
*X=Color Number*
*Y=Return Code*
*Z=Temp*
*Output: Return Code*
*Begin*
*Step1. Y = X AND 0x0A*
*Step2. Z = (X AND 0x01) SHR 2*
*Step3. Y = Y OR Z*
*Step4. Z = (X AND 0x04) SHL 2*
*Step5. Y = Y OR Z*
*End*

Figure 1: Conversion Algorithm

4-bit color bitmap can be converted to 4-shade grey scale image without using the conventional color to grey image conversion formula i.e.

$$Grey = 0.3 * R + 0.11 * B + 0.59 * G \qquad \ldots\ldots (1)$$

The result of the same is shown in Table 2. Return Code 0, 7, 8, 15 remains unchanged. Rest codes will be changed as per the following algorithm (see figure 2).

Table 2: Return Code to Grey Scale Code

| Colors | RC | RC Binary | | | | After 3rd SHR | | | | GC |
|---|---|---|---|---|---|---|---|---|---|---|
| | | I | R | G | B | $X_4$ | $X_3$ | $X_2$ | $X_1$ | |
| Black | 0 | 0 | 0 | 0 | 0 | 0 | 0 | 0 | 0 | 0 |
| Red | 4 | 0 | 1 | 0 | 0 | 0 | 0 | 0 | 0 | 0 |
| Green | 2 | 0 | 0 | 1 | 0 | 0 | 0 | 0 | 0 | 0 |
| Brown | 6 | 0 | 1 | 1 | 0 | 0 | 0 | 0 | 0 | 0 |
| Blue | 1 | 0 | 0 | 0 | 1 | 1 | 0 | 0 | 0 | 8 |
| Magenta | 5 | 0 | 1 | 0 | 1 | 1 | 0 | 0 | 0 | 8 |
| Cyan | 3 | 0 | 0 | 1 | 1 | 1 | 0 | 0 | 0 | 8 |
| Light Gray | 7 | 0 | 1 | 1 | 1 | 0 | 1 | 1 | 1 | 7 |
| Dark Gray | 8 | 1 | 0 | 0 | 0 | 1 | 0 | 0 | 0 | 8 |
| Light Red | 12 | 1 | 1 | 0 | 0 | 0 | 1 | 1 | 1 | 7 |
| Light Green | 10 | 1 | 0 | 1 | 0 | 0 | 1 | 1 | 1 | 7 |
| Yellow | 14 | 1 | 1 | 1 | 0 | 0 | 1 | 1 | 1 | 7 |
| Light Blue | 9 | 1 | 0 | 0 | 1 | 1 | 1 | 1 | 1 | 15 |
| Light Magenta | 13 | 1 | 1 | 0 | 1 | 1 | 1 | 1 | 1 | 15 |
| Light Cyan | 11 | 1 | 0 | 1 | 1 | 1 | 1 | 1 | 1 | 15 |
| White | 15 | 1 | 1 | 1 | 1 | 1 | 1 | 1 | 1 | 15 |

*RC = Return Code, GC = Grey Code, I=Intensity, R=Red, G=Green, B=Blue

**Algorithm 2** *Color_Grey*
*Input: Return Code*
*Output: 4-shade Grey code*
*Begin*
    Step1.   Temp = RC
    Step2.   A = (Temp AND 0x04) SHL 3
    // Copy the Intensity bit (0 or 1) to variable A
    Step3:   Loop i = 1 to 3
        RC = RC SHL 1 // RC left shift
        RC = RC OR A
        // Intensity bit is copied to LSB
        End Loop
*End*

Figure 2: Color Change Algorithm

After execution of the aforesaid algorithm, a 4-bit binary code will be generated, out of which the 3$^{rd}$ and 4$^{th}$ bit will be considered (see Table 3) and 1$^{st}$ and 2$^{nd}$ bit will be discarded to generate the 2-bit binary code (grey code) i.e. 00, 01, 10, 11 or 4-shade grey scale i.e. 0000, 0111, 1000, 1111.

Table 3: 3$^{rd}$ and 4$^{th}$ Bit of Gray Scale Code

| Colors | 4-bit GC | Binary | 4$^{th}$ & 3$^{rd}$ Bit |
|---|---|---|---|
| Black | 0 | **00**00 | 00 |
| Light Gray | 7 | **01**11 | 01 |
| Dark Gray | 8 | **10**00 | 10 |
| White | 15 | **11**11 | 11 |

Using the aforesaid logic any 4-bit color bitmap image can be converted to 4-shade greyscale image without conventional method. Figure 3 depicts the high level flow of above stated procedure.

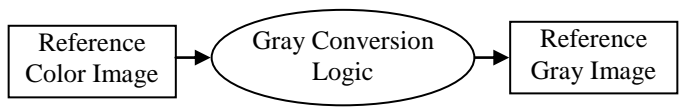

Figure 3: Color to Gray Scale Image

### *3.2 Hiding Information*

Using this grey scale reference image any text can be hidden. Single character of a text can be represented by 8-bit. The 8-bit character can be split into 4X2 bit information. Each 2-bit of information must be 00, 01, 10 or 11. These combinations are present in the grey scale reference image once or several occasions. If we use the image as reference, we have to store only the coordinate of the occurrences to a data file to cipher the text. We can use the random occurrence to make the system more robust. The field of the data file will be occurrence, x coordinate and y coordinate. The following figure 4(a) and algorithm in figure 4(b) describes the process in details:

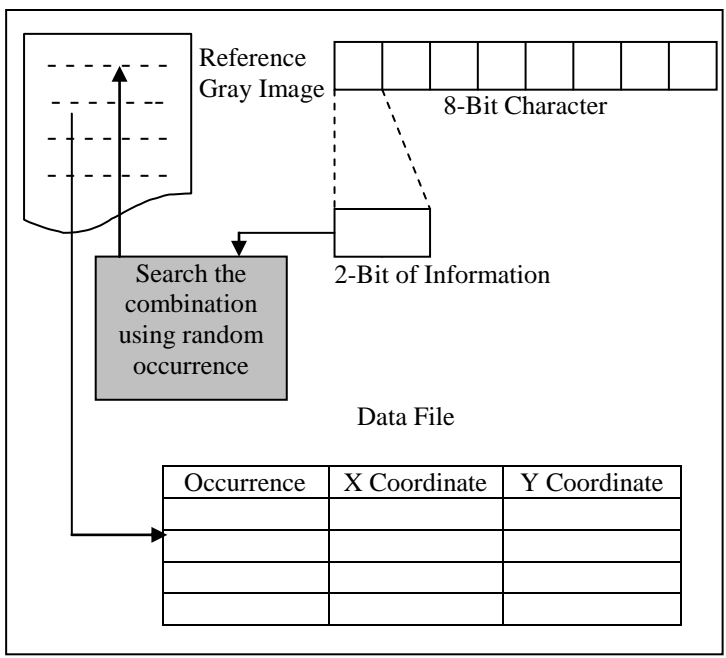
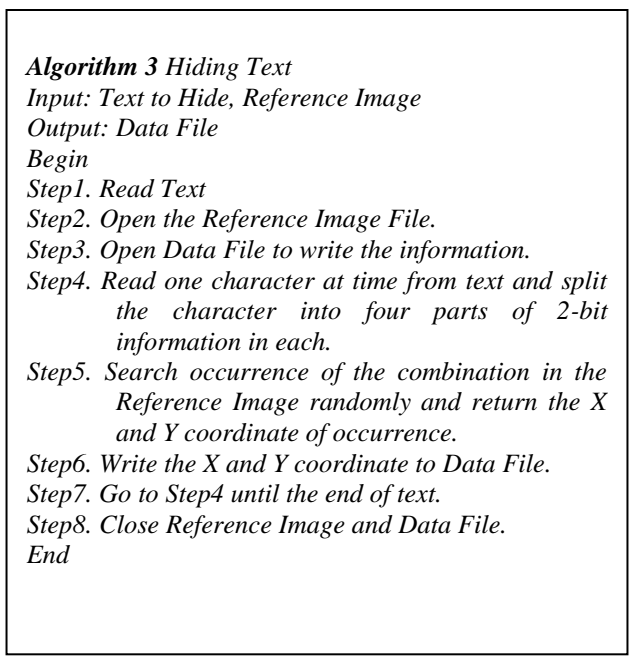

Figure 4(a): Ciphering Technique          Figure 4(b): Algorithm for Hiding Text

### *3.3 Transmission and Unhiding*

If the reference image and the data file are transmitted through network separately, we can achieve the effect of Steganography. Here the image is not at all distorted because said image is only used for referencing. Any huge amount of text material can be hidden using a very small image. Decoding of information is also very easy. The coordinate of the image has to read from the data file and collect the bit information from the reference image to reconstruct the text information. The transition process and the algorithm are depicted in figure 5(a) and 5(b) respectively.

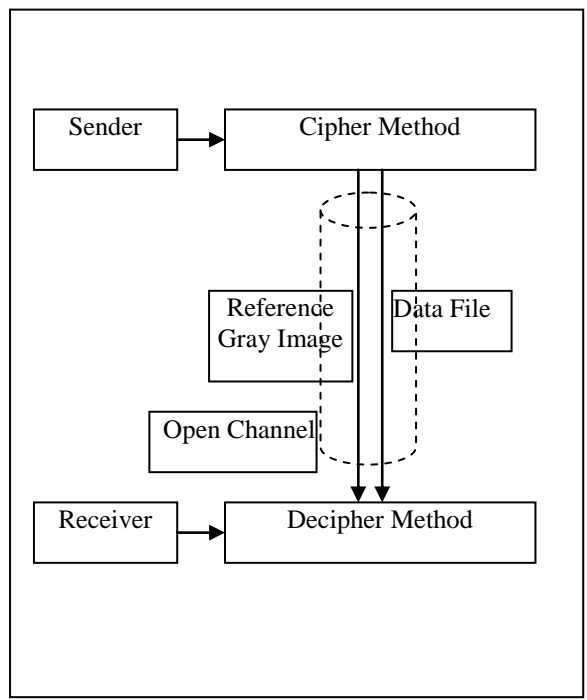

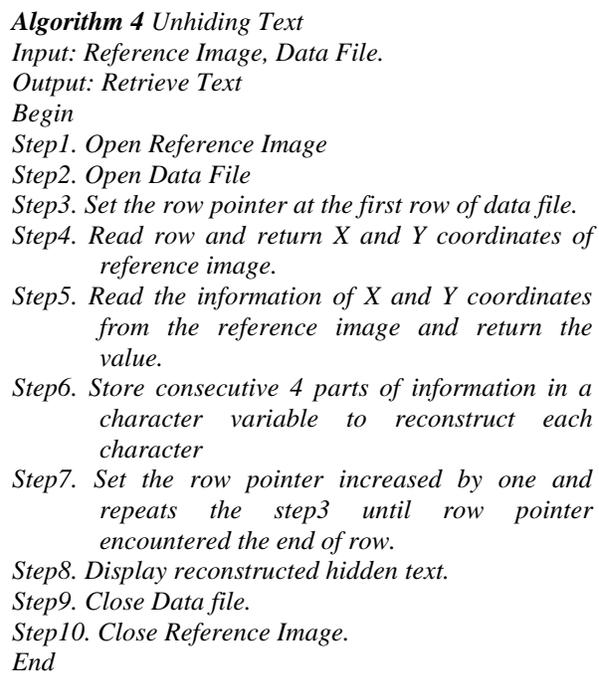

*Algorithm 4* Unhiding Text
Input: Reference Image, Data File.
Output: Retrieve Text
Begin
*Step1. Open Reference Image*
*Step2. Open Data File*
*Step3. Set the row pointer at the first row of data file.*
*Step4. Read row and return X and Y coordinates of reference image.*
*Step5. Read the information of X and Y coordinates from the reference image and return the value.*
*Step6. Store consecutive 4 parts of information in a character variable to reconstruct each character*
*Step7. Set the row pointer increased by one and repeats the step3 until row pointer encountered the end of row.*
*Step8. Display reconstructed hidden text.*
*Step9. Close Data file.*
*Step10. Close Reference Image.*
End

Figure 5(a): Transition Method                Figure 5(b): Algorithm for Unhiding Text

Since, deciphering the text is not possible by intercepting the image or data file separately. So, it is more secure

### 4. Conclusions

In this paper we have explained the basic mechanism of our proposed model and it is an alternative approach of steganography. It is not pure steganographic technique but the effect is same with some additional advantage. First advantage is the data file and reference image is going through the open channel separately. The basic result is interception of any one cannot provide desired objective. Second advantage is that any amount of data can be transmitted using the method because it is not depending on the size of image. Final advantage, the said method is not affecting the image. There is no change of quality and color change of reference image. It is most vital achievement of method.

The algorithm time complexity is simple and always proportional to $O(n)$. The performance of hiding algorithm is totally depending on the length of text to hide and size of image. Similarly Unhiding algorithm is reverse process of previous one and complexity character is same. At end, this can be said that the aforesaid method may be improved, instead of text small image may be hiding, cryptography may be used or much improvement in this field may be incorporated in future. Lastly it is expected by the authors that any kind of future endeavours in this field will definitely route it a path to design a secure system using the proposed algorithm for both Internet and Mobile Communication Technology.